\documentclass[conference]{IEEEtran}
\usepackage{graphicx}
\usepackage{caption}
\usepackage{subcaption}

\usepackage{longtable}
\usepackage{stfloats}
\usepackage{graphicx}% http://ctan.org/pkg/graphicx

\usepackage{booktabs}
\usepackage{multirow}

\usepackage{ifpdf}
\usepackage[T1]{fontenc}
\usepackage{amssymb}
\usepackage{cite}
\pdfoutput=1

\ifCLASSINFOpdf
   \usepackage{graphicx}
	\usepackage{epstopdf}
   \graphicspath{{images/}}   
   \graphicspath{{../pdf/}{../jpeg/}{../eps/}}
   \DeclareGraphicsExtensions{.pdf,.jpeg,.png,.eps}
\else
   \usepackage[dvipdfm]{graphicx}
   \usepackage{bmpsize}
   \graphicspath{{../eps/}}
   \DeclareGraphicsExtensions{.eps}
\fi
\graphicspath{{images/}}

\usepackage[cmex10]{amsmath}
\usepackage{multirow}

\usepackage{multirow}

\usepackage{booktabs}
\usepackage{multicol}

\usepackage[shortcuts,acronym, nomain]{glossaries}

\makeglossaries % generates the acronym list

\newacronym{rnn}{RNN}{Recurrent Neural Network}
\newacronym{awgn}{AWGN}{Additive White Gaussian Noise}
\newacronym{gru}{GRU}{Gated Recurrent Unit}
\newacronym{lstm}{LSTM}{Long-Short Term Memory}
\newacronym{snr}{SNR}{Signal to Noise Ratio}
\newacronym{urllc}{URLLC}{Ultra Reliable Low Latency Communication}
\newacronym{dl}{DL}{Deep Learning}
\newacronym{ml}{ML}{Machine Learning}
\newacronym{qos}{QoS}{Quality of Service}
\newacronym{mmtc}{MMTC}{Massive Machine Type Communications}
\newacronym{iot}{IoT}{Internet of Things}
\newacronym{cnn}{CNN}{Convolutional Neural Network}
\newacronym{gan}{GAN}{Generative Adversarial Network}
\newacronym{dbn}{DBN}{Deep Belief Network}
\newacronym{phy}{PHY}{Physical Layer}
\newacronym{ann}{ANN}{Artificial Neural Network}
\newacronym{mlp}{MLP}{Multi Layer Perceptron}
\newacronym{ldpc}{LDPC}{Low-Density Parity-Check}
\newacronym{ber}{BER}{Bit Error Rate}
\newacronym{gpu}{GPU}{Graphical Processing Unit}
\newacronym{rsc}{RSC}{Recursive Systematic Convolutional}
\newacronym{map}{MAP}{Maximum A Posteriori}
\newacronym{adam}{ADAM}{Adaptive Moment Estimation}
\newacronym{qpsk}{QPSK}{Quadrature Phase Shift Keying}
\newacronym{qam}{QAM}{Quadrature Amplitude Modulation}

\begin{document}

\title{Performance Analysis of Deep Learning based on Recurrent Neural Networks for Channel Coding}
\author{\IEEEauthorblockN{Raja Sattiraju, Andreas Weinand and Hans D. Schotten}
\IEEEauthorblockA{Chair for Wireless Communication \& Navigation \\
University of Kaiserslautern\\
\{sattiraju, weinand, schotten\}@eit.uni-kl.de}}
\maketitle

\begin{abstract}
Channel Coding has been one of the central disciplines driving the success stories of current generation LTE systems and beyond. In particular, turbo codes are mostly used for cellular and other applications where a reliable data transfer is required for latency-constrained communication in the presence of data-corrupting noise. However, the decoding algorithm for turbo codes is computationally intensive and thereby limiting its applicability in hand-held devices. In this paper, we study the feasibility of using \ac{dl} architectures based on \acp{rnn} for encoding and decoding of turbo codes. In this regard, we simulate and use data from various stages of the transmission chain (turbo encoder output, \ac{awgn} channel output, demodulator output) to train our proposed \ac{rnn} architecture and compare its performance to the conventional turbo encoder/decoder algorithms. Simulation results show, that the proposed \ac{rnn} model outperforms the decoding performance of a conventional turbo decoder at low \ac{snr} regions.

\end{abstract}

\section{Introduction}

{\let\thefootnote\relax\footnote{This is a preprint, the full paper will be published in Proceedings of 12th International Conference on Advanced Networks and Telecommunications Systems Conference (IEEE ANTS 2018), \copyright 2018 IEEE. Personal use of this material is permitted. However, permission to use this material for any other purposes must be obtained from the IEEE by sending a request to pubs-permissions@ieee.org.}}

The 5G wireless networks are a major transformational force that is propelled by the explosion of radio devices along with advanced and not yet supported applications and services spanning across multiple domains such as industrial and vehicular communications. Beyond the need for high data rates, a factor that has been the primary driver for the evolution of wireless networks until 4G, 5G networks should also be able to support diverse \ac{qos} requirements such as \ac{urllc}, \ac{mmtc} etc. Added to this, the \ac{iot} ecosystem would necessitate collecting short packets of periodic data in real time thus leading to substantial traffic on the uplink channel. These new applications also bring with them complex scenarios with unknown channel models, high speed and accurate processing requirements \cite{Wang2017}; thereby challenging conventional communication system algorithms and hence paving the way for a paradigm shift in the design of communication systems.\

Driven by this demand, \ac{ml}, which has shown substantial promises during recent years\cite{Goodfellow} is extensively studied by researchers in order to assess its applicability to wireless networks. \acf{dl}, which belongs to a class of \ac{ml} algorithms, that uses multiple layers of non linear processing units stacked on top of each other. Each successive layer uses the output of the previous layer as input. Such DL architectures are especially suitable for designing auto encoders that aim to find a low-dimensional representation of its input at some intermediate layer that allows reconstruction at the output with minimal error. DL architectures such as \acp{cnn}, \acp{rnn}, \acp{gan}, \acp{dbn} have been applied to various domains such as computer vision, natural language processing, social network filtering, drug design etc. where they have produced results comparable to and in some cases superior to human experts.

Initially applied to upper layers, such as resource management\cite{Challita2017}, link adaptation\cite{Daniels2010,Pulliyakode2017}, obstacle detection \cite{Sattiraju2017a} and localization\cite{Vieira2017}, \ac{ml} has recently found applications also at the \ac{phy} functions such as channel coding\cite{Ortuno1992,Bruck1989}, modulation recognition\cite{Nandi1997,OShea2017}, physical layer security \cite{Weinand2017}, and channel estimation and equalization\cite{Wen2015,Chen1990} etc.\

The first attempts for using \acp{ann} for decoding turbo codes were presented in \cite{Annauth} where the author proposes an \ac{ann} based on \acp{mlp}. With the advent of training techniques such as layer-by-layer unsupervised pre-training followed by gradient descent fine-tuning and back propagation, the interest for using \acp{ann} for channel coding is renewed. Different ideas around the use of \acp{ann} for decoding emerged in the 1990s with works such as \cite{Caid, DiStefano, Abdelbaki} for decoding block and hamming codes. Subsequently, \acp{ann} were used for decoding convolutional codes in \cite{Xiao-AnWang1996, Kao2009}. In \cite{Wei}, the author used \acp{mlp} to generate \ac{ldpc} codes. More recent works such as \cite{Gruber2017} use the more advanced deep \acp{ann} for decoding structured polar codes.

%\subsection{Contribution \& Motivation of this work}
In this work, we investigate DL architectures to design and analyse an \ac{ann} for turbo coding and decoding operations that are typically performed at the \ac{phy}. Specifically, we use the turbo encoder and decoder variant specified for LTE \cite{TSGR2017c}. To this end, we frame the encoding and decoding operations as a supervised learning problem and use a \ac{rnn} architecture to autoencode-decode the data and compare its performance in terms of \ac{ber} to the legacy LTE turbo encoding/decoding blocks. Our motivation is two fold

\begin{enumerate}
	\item[i.] Traditional signal processing is done by logically separated blocks that are independently optimized to recover the data signal from imperfect channels. Although this approach is perfected over many years, it may not achieve the optimal end-to-end performance. A well learned DL encoder/decoder is by default optimized for end-to-end performance.
	\item[ii.] \acp{ann} are shown to be universal function approximators\cite{Hornik1989} and are known to be Turing complete\cite{Siegelmann1995}. Their execution can be highly parallelized on distributed memory architectures such as \acp{gpu} that have high energy efficiency and computational throughput. Hence, these learned algorithms could be executed faster and at lower energy cost than traditional signal processing blocks.
\end{enumerate}

This work is organized as follows. Section II presents a brief overview of the turbo encoding and decoding operations used in LTE. A general introduction to \ac{dl} and \acp{rnn} is presented in Section III. Section IV explains the problem formulation, data preparation, model architecture and the training operation with the corresponding numerical results presented in Section V. Section VI concludes the paper and presents some discussions for future work. 

\section{Turbo Encoder Architecture and Interfaces}
\begin{figure}
	\centering 
	\includegraphics[width=0.48\textwidth]{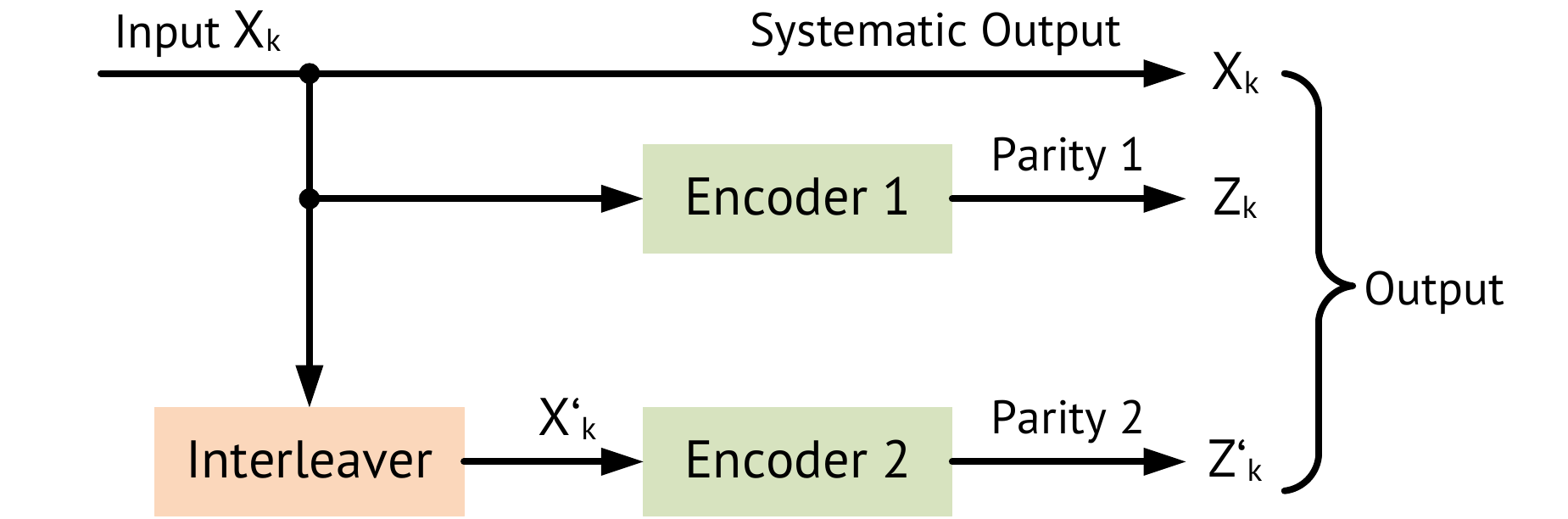}
	\caption{Turbo Encoder Architecture}
	\label{fig:turboencoder}
\end{figure}

In general sense, a turbo encoder consists of two encoders (referred to as constituent encoders) separated by an interleaver. The encoders are normally identical and the interleaver is used to scramble the bits before being fed to the second encoder. Thus the encoder outputs are different from each other. LTE uses two 8-state identical \ac{rsc} encoders that are concatenated in parallel and separated by an internal interleaver \cite{TSGR2017c} as shown in Fig. \ref{fig:turboencoder}.

The transfer function of each constituent encoder is given as
\[G(D) = \begin{bmatrix}
1, \frac{g_1(D)}{g_0(D)} 
\end{bmatrix}  \]

where $g_0(D) = 1 + D^2 + D^3$ and $g_1(D)=1+D+D^3$

For a given input bit stream $X_0, X_1,...,X_{K-1}$ of length $K$, the output of the turbo encoder is given as 
\[X_0, Z_0, Z'_0, X_1, Z_1, Z'_1,...,X_{K-1}, Z_{K-1}, Z'_{K-1}  \]

where
\begin{enumerate}
	\item Bits $X_0,X_1,...,X_{K-1}$ are the systematic bits as well as the input to the first constituent encoder and the internal interleaver
	\item Bits $Z_0,Z_1,...,Z_{K-1}$ and $Z'_0,Z'_1,...,Z'_{K-1}$ are outputs from the first an second constituent encoders
\end{enumerate}

As can be seen from the Fig. \ref{fig:turboencoder}, bits $X'_0,X'_1,...X'_{K-1}$ are outputs from the internal interleaver (and the input to the second constituent encoder) for which the relationship between input and output bits is given as
\[ X'_i = X_{\pi(t)},\: i=0,1,...,K-1 \]
where the relationship between the output index $i$ and the input index, $\pi(i)$ satisfies the following quadratic form
\[ \pi(i) = (f_1 i + f_2i^2) \] 
where parameters $f_1$ and $f_2$ depend on the block size $K$. The valid block lengths along with their corresponding $f_1$, $f_2$ values are given in the 3GPP specification \cite{TSGR2017c}.

Each bit stream is trellis terminated, by taking the tail bits from the shift register after encoding and padding them to the stream bits. This is done so as to reset the encoder state to zero after every encoding operation. Hence, for any given $k$ bits, the output length of the turbo encoder is $3*k + 3*m$ where $m$ is the memory size in the shift registers. For LTE variant, the size of $m$ is 4 and hence 4 tail bits are added to each stream totalling 12 bits. 

\begin{figure}
	\centering 
	\includegraphics[width=0.48\textwidth]{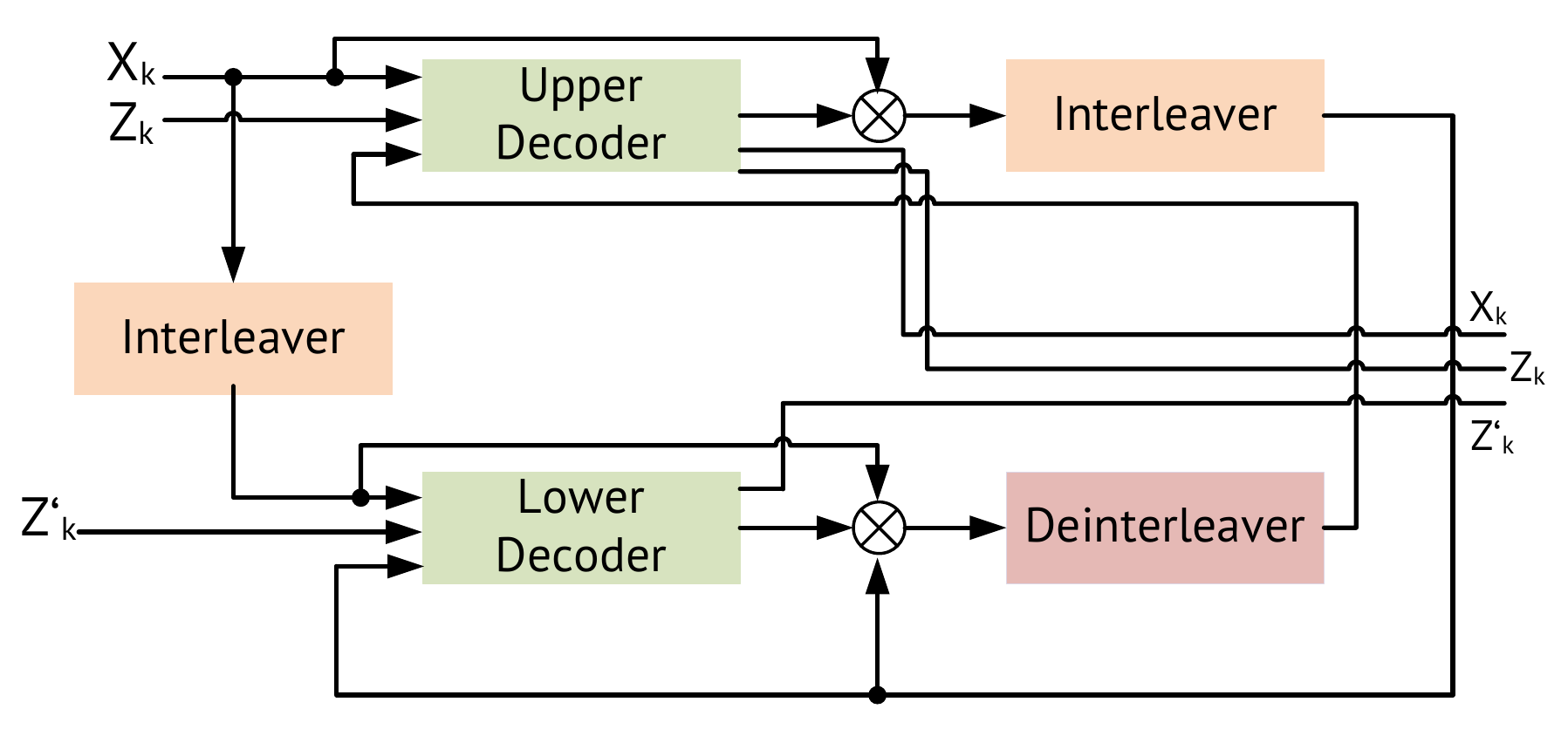}
	\caption{Turbo Decoder Architecture}
	\label{fig:turbodecoder}
\end{figure} 

At the receiver end, the turbo decoder consists of two single soft-in soft-out (SISO) decoders that work iteratively. As seen from Fig. \ref{fig:turbodecoder}, the output of the upper decoder feeds into the lower decoder to form a turbo decoding iteration. Interleaver and deinterleaver blocks re-order data in this process. Two decoding algorithms based on \ac{map}, namely \textit{LogMAP} and \textit{MaxLogMAP} are used for the decoding process.

\section{\acl{dl} for Turbo Encoding and Decoding}
Any \ac{ml} algorithm learns the execution of a particular task $T$, maintaining a specific performance metric $M$, based on exploiting its experience $E$\cite{Jiang2017}. \ac{ml} algorithms can be classified into Supervised and Unsupervised depending on the presence/absence of labeled samples in the input dataset. \ac{dl} is an emerging algorithm belonging to the class of \acp{ann} which consists of multiple stacked layers with each layer consisting of an arbitrary number of neurons. At each neuron, all of its weighted inputs are added up along with a bias and the result is propagated to the next layer through a nonlinear activation function. Each layer $i$ with $n_i$ inputs and $m_i$ outputs performs a mapping $f^{(i)}:\mathbb{R}^{n_i} \mapsto \mathbb{R}^{m_i}$. Denoting $r_o$ as output and $r_i$ as input of the ANN, the input-output mapping is defined by a chain of functions depending on the parameter set $\theta$ (weights and biases) by
\[ 
	r_o = f(r_i;\theta) = f^{(L-1)}\left ( f^{(L-2)}\left ( ...\left ( f^{(0)}(r_i) \right ) \right ) \right )
 \]
 
where $L$ is the number of layers in the \ac{ann} (also referred to as \textit{depth}). In order to find the optimal weights of the \ac{ann}, a training operation with a specific loss function is defined. During training, the \ac{ann} adjusts its weights to minimize the loss function over the training set by means of gradient descent optimization methods and the backpropagation algorithm.\

\acp{ann} can be broadly classified into two types - feed forward networks  where the information moves only in the forward direction from input nodes, through the hidden nodes and to the output nodes. Examples of such ANN architectures include a simple perceptron, \acp{cnn}. These are suitable for learning unconnected inputs such as image / video processing. In this paper, we are dealing with turbo encoding and decoding operations that exhibit temporal dependencies between input sequences. Hence, our discussion is limited to \acp{rnn} that are better suited to learning connected inputs. Unlike traditional feed forward networks, \acp{rnn} have an internal hidden state (memory) whose activation at each time is dependent on that of the previous time and are hence suitable to process connected input sequences. Formally, given a sequence $(x_1, x_2,...,x_T)$, the \ac{rnn} updates its recurrent hidden state $h_t$ as follows

\[ h_t = g(Wx_t + Uh_{t-1}) \]
where $g$ is a smooth, bounded non-linear function (e.g., sigmoid function), and $W$, $U$ are parameters of the network.\

Out of many available \acp{rnn}, two variants stand out. The first is based on \ac{lstm} \cite{Hochreiter1997a} and the second one is based on the more recent \ac{gru}\cite{Cho2014}. The key difference between them is that \ac{lstm} uses three gates (namely input, output and forget gates) whereas \ac{gru} uses two gates (reset and update gates). The flow of information is similar for both the architectures except that \ac{gru} does not use a memory unit and exposes the full hidden content without any control. The performance of \ac{gru} is almost on par with \ac{lstm} albeit at a lower computational complexity. It is because of these reasons that we select \acp{gru} as the underlying units for our proposed \ac{rnn}.

\subsection{\acf{gru}}
\ac{gru} was proposed in \cite{Cho2014} to make each unit to adaptively capture dependencies of different time scales. The activation $h_t$ of a \ac{gru} at time $t$ is a linear interpolation between the previous activation $h_{t-1}$ and the target activation $\tilde{h}_t$
\[ h_t = (1-z_t)h_{t-1} + z_t\tilde{h}_t \]

where the update gate $z_t$ decides how much the unit updates it activation and is computed as
\[ z_t = \tanh(Wx_t + U(r_t \odot h_{t-1}))^j \]

where $\odot$ is an elementwise multiplication and  $r_t$ is a set of reset gates that are computed similarly to the update gates as

\[ r_t = \sigma(W_rx_t + U_rh_{t-1})^j \]

\subsection{Activation Function}
The activation used in all the models is a sigmoid function to induce non-linearity to the outputs and also due to its good binary classification capability. It is a special case of a logistic function and is defined by the formula
\[ S(x) = \frac{1}{1 + e^{-x}} \]

\subsection{Cost Funtion}
The cost or loss function is used to evaluate the performance of a given network and update the weights accordingly. Due to the binary nature of our activation function, we selected binary cross-entropy as the loss function which can be calculated as
\[ f(l) = -(y\log(p))+(1-y)\log(1-p) \]

where $y$ is the binary indicator (0 or 1) if the output value is equal to the true value and $p$ is the predicted probability of an observation being either 0 or 1.

\subsection{Optimizer}
For all the models, \ac{adam} optimizer was used. It computes the learning rates by calculating an exponentially decaying average of past gradients $m_t$ in addition to past squared gradients $v_t$ as follows\cite{Kingma2014}

\begin{equation*} 
\begin{split}
m_t & = \beta_1m_{t-1}+(1-\beta_1)g_t \\
v_t & = \beta_2v_{t-1}+(1-\beta_2)g^2_t
\end{split}
\end{equation*}

These values are then used to update the weights according to following rule
\[ \theta_{t+1} = \theta_t - \frac{\eta}{\sqrt{v_t} + \epsilon}m_t \]

\section{Simulation Framework}
\begin{figure}[h]
	\centering 
	\includegraphics[width=0.48\textwidth]{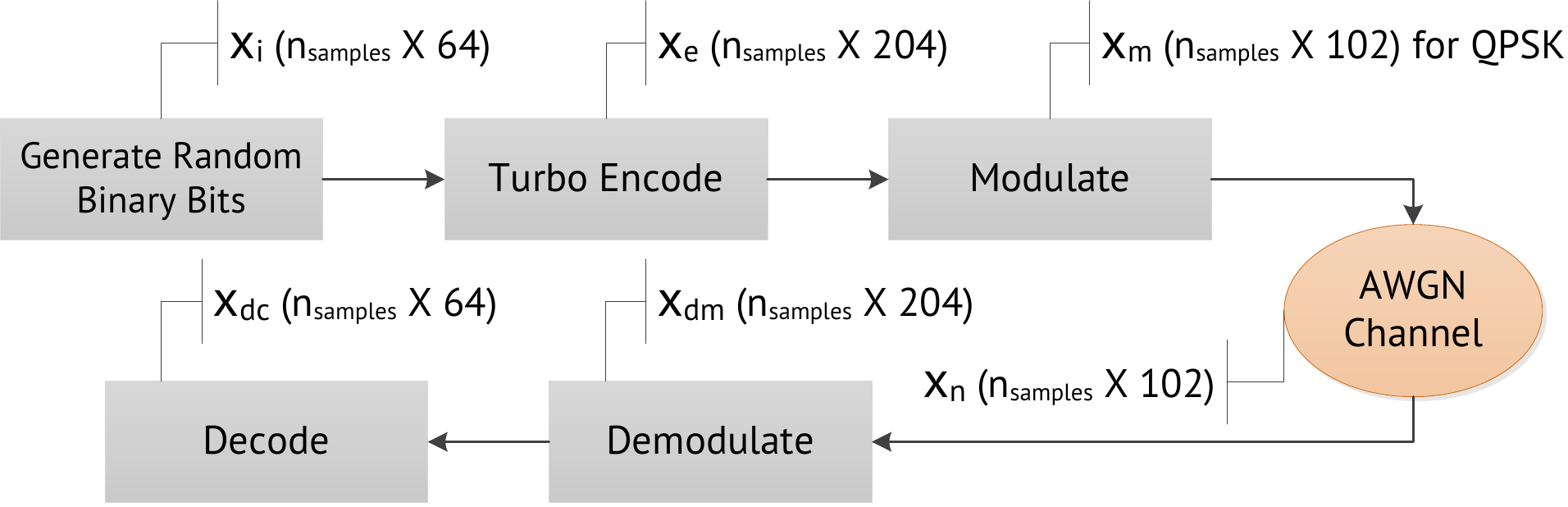}
	\caption{Simulation Method}
	\label{fig:simMethod}
\end{figure}

In this work, we investigate the performance of \acp{rnn} based on \acp{gru} for encoding and decoding of turbo codes. In order to train the model, data was generated using the communications system toolbox in Matlab and the sequence of operations are highlighted in Fig. \ref{fig:simMethod} 

\subsection{Problem Formulation, Data Generation \& Preparation}

A total of four autoencoding problems have been formulated - one for encoding and the remaining three for decoding as follows.

\subsubsection{Turbo Encoding}
A series of binary bits are given as input to the \ac{rnn} whose goal is to encode them by adding redundancy and output the turbo encoded bits. 10000 packets of size 64 bits were generated i.e., $\text{size}(x_i) = (10000,64)$ as input and we used the LTE variant of turbo encoder to encode the data. Because of the trellis termination, tail bits are added to each data stream thus creating the resulting encoded data $x_e$ of size $(10000,204)$. No \ac{awgn} noise is considered for this scheme.

\subsubsection{Turbo Decoding}
For the turbo decoding, three data generation approaches were considered

\begin{enumerate}
	\item[i.] Reversing the $x_i$ and $x_e$ obtained from the previous step and feeding the encoded data $x_e$ to the \ac{rnn} to obtain the decoded bits $x_i$. This is the simplest case without considering any noise and modulation.
	\item[ii.] Using the demodulated soft bits $x_{dm}$ which is fed to the \ac{rnn} in-order to obtain the decoded bits $x_{i}$. Data is generated for different \acp{snr} in range [-2,2) totalling 11*10000 = 110000 samples.
	\item[iii.] Using the noise affected data $x_n$ which is of complex data type directly without demodulation and feeding it as input to the \ac{rnn} to obtain the decoded bits $x_i$. Similar to the second approach, data is generated for different \acp{snr} in the range [-2,2) totalling 110000 samples.
\end{enumerate}

For each auto-encoding problem, we shuffle and split the input data into training and testing datasets in the ratio of 30\%-70\% respectively. No scaling/preprocessing has been used.

\subsection{RNN Model, Training \& Validation}

\begin{table}[htb!]
	\caption{Model Architecture}
	\label{tab:modelarch}
	\begin{subtable}{0.5\textwidth}
		\centering
		\caption{Input Shapes}
		\label{tab:inputshapes}
		\begin{tabular}{@{}lll@{}}
			\toprule
			Problem            & Input Shape             & Output Shape    \\ \midrule
			Turbo Encoding     & ($N_s$, 1, 64)       & ($N_{s}$, 204) \\
			Turbo Decoding - 1 & ($N_{s}$, 1, 204)      & ($N_{s}$, 64)  \\
			Turbo Decoding - 2 & ($N_{s}$, 1, 204)       & ($N_{s}$, 64)  \\
			Turbo Decoding - 3 & ($N_{s}$, 2, 204 / m*) & ($N_{s}$, 64)  \\ \midrule
			\multicolumn{3}{l}{$N_s$ is batch size}                       \\  
			\multicolumn{3}{l}{*m = 2,4,6 for QPSK, 16QAM and 64QAM}      \\ \bottomrule   
		\end{tabular}
	\end{subtable}
	
	\bigskip
	
	\begin{subtable}{0.5\textwidth}
		\centering
		\caption{Layers}
		\label{tab:modellayers}
		\begin{tabular}{@{}lll@{}}
			\toprule
			Layer                     & Shape           & Parameters   \\ \midrule
			Input                     & Refer to \ref{tab:inputshapes}    & None         \\
			\ac{gru} - 1                   & ($N_{s}$,1,800)   & 2078400      \\
			Batch Normalization - 1   & ($N_{s}$,1,800)   & 3200         \\
			\ac{gru} - 2                   & ($N_{s}$,800)     & 3844800      \\
			Batch Normalization - 2   & ($N_{s}$,800)     & 3200         \\
			Dense                     & Refer to \ref{tab:inputshapes}     & Variable       \\ \midrule
			\multicolumn{3}{l}{Total Trainable Parameters $\approx$ 6,000,000} \\ \bottomrule
		\end{tabular}
	\end{subtable} 
\end{table}
The model used in this work is similar to the one presented in \cite{Kim2018} and can be seen in Table \ref{tab:modellayers}. It consists of two layers of bidirectional \acp{gru} with each layer followed by a batch normalization layer. The output layer is a single fully connected sigmoid unit. The bi-directionality is intended to support recursion in both forward pass and backward pass through the received sequence. It has to be noted that the same model is used for all the autoencoding problems by modifying the input shape (Table \ref{tab:inputshapes})

The training set is further split into model training and validation datasets in the ratio of 90\%-10\%. The model is trained on the training dataset on an Nvidea \ac{gpu} for 30 epochs (an \textit{epoch} is one pass over the entire dataset) and validated on the validation set. Finally, the model is applied on the testing dataset.

\section{Results}

For problem sets 1 and 2, the training, validation and testing accuracies for a single \ac{snr} point were used for performance evaluation since there is no noise added to the input data. For problem set 3 and 4, the testing \ac{ber} $(1-\text{testing accuracy})$ per \ac{snr} point was used.

\subsection{Turbo Encoding and Decoding on data with no noise}

\begin{figure}[h!]
	\centering 
	\includegraphics[width=0.48\textwidth]{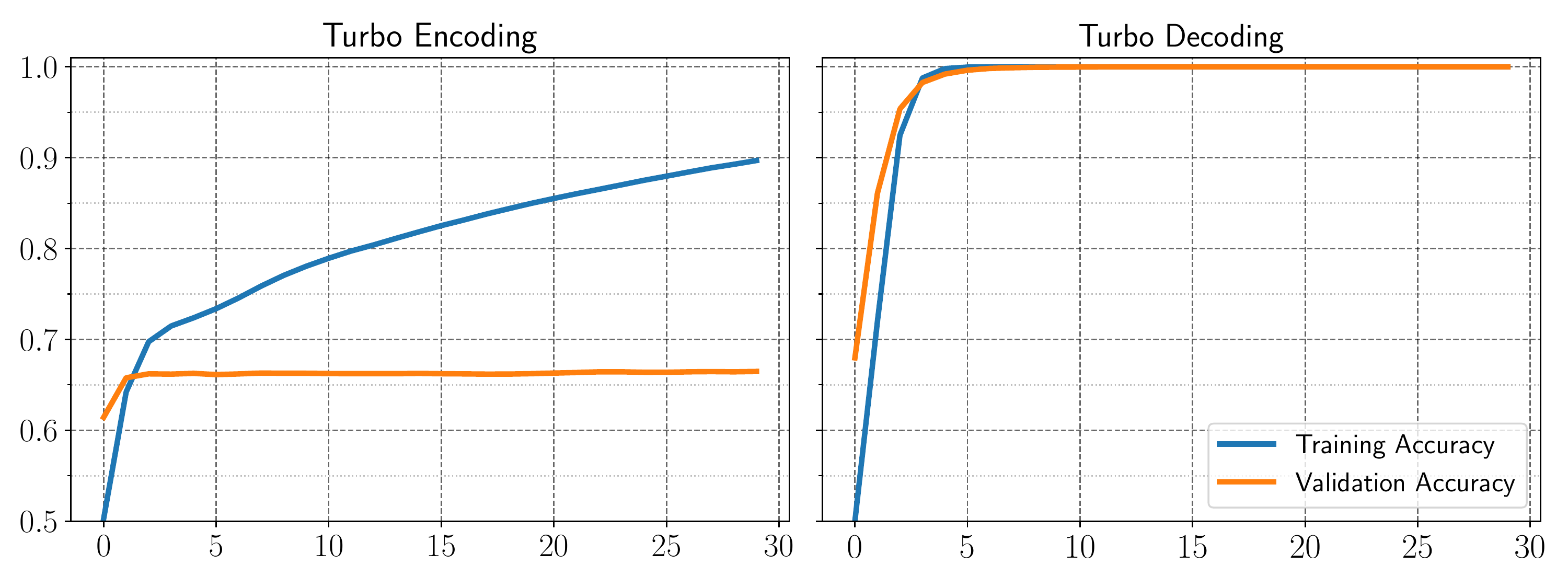}
	\caption{Training \& Validation Accuracy - No Noise}
	\label{fig:withoutNoise}
\end{figure} 

\begin{figure}[h!]
	\centering
	\begin{subfigure}[b]{0.49\textwidth}
		\includegraphics[width=1\linewidth]{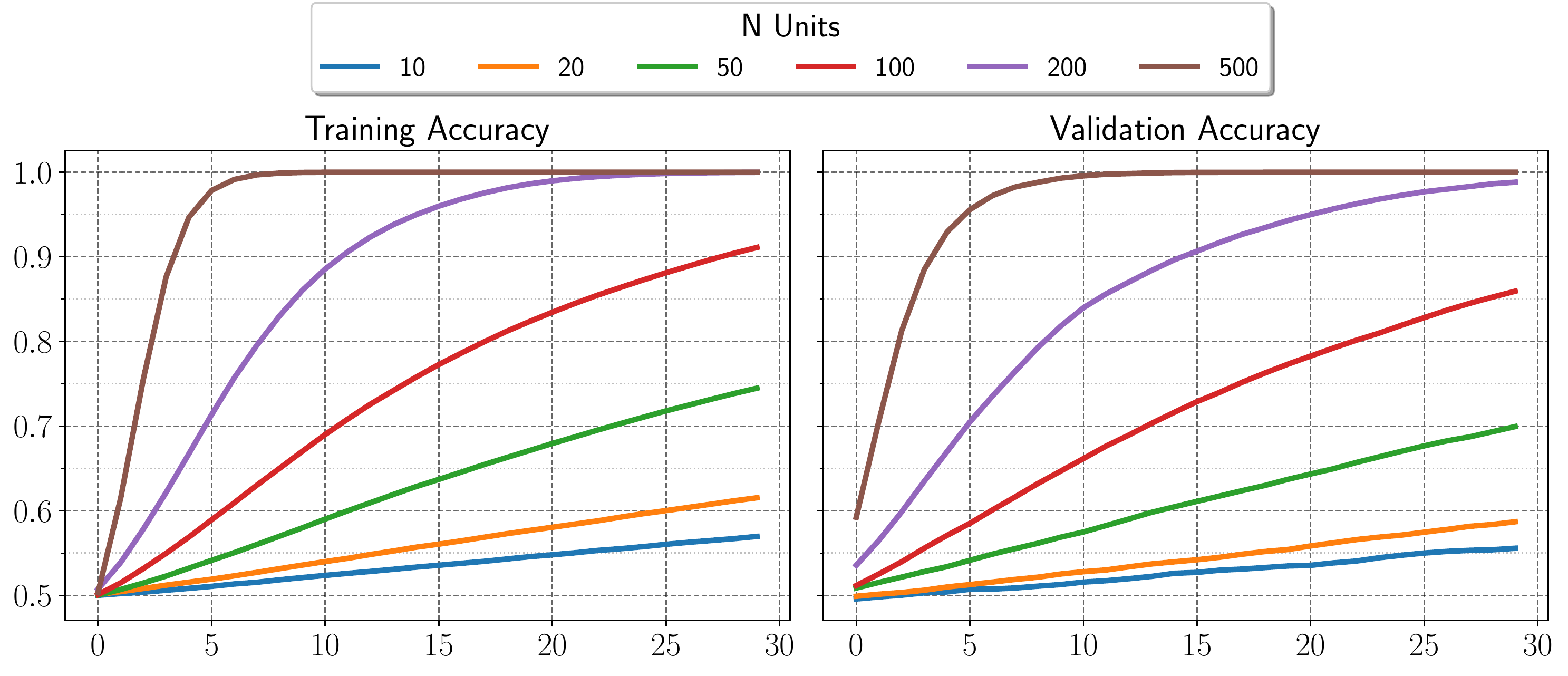}
		\caption{Effect of Number of \ac{gru} Units on Accuracy}
		\label{fig:Ng1} 
	\end{subfigure}
	
	\begin{subfigure}[b]{0.49\textwidth}
		\includegraphics[width=1\linewidth]{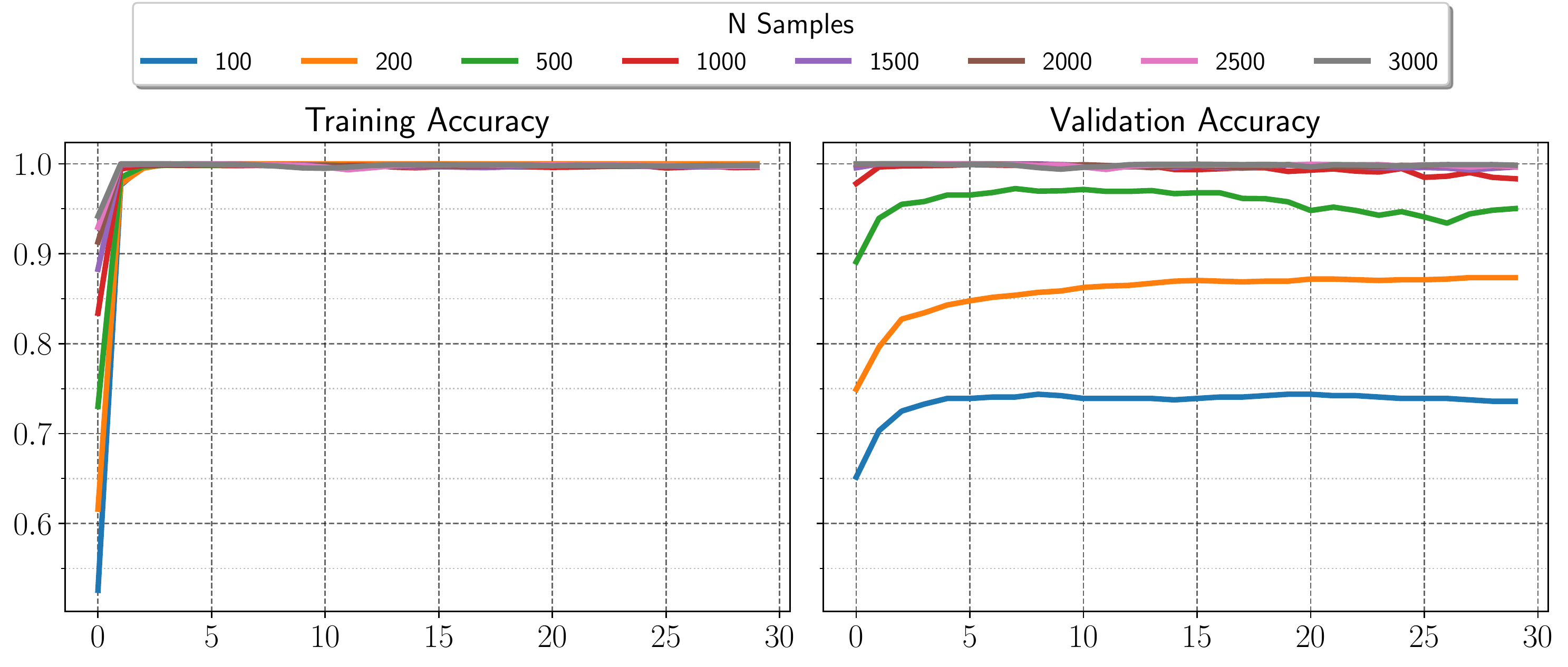}
		\caption{Effect of Number of training samples on Accuracy }
		\label{fig:Ng2}
	\end{subfigure}
	\caption{Model Selection \& Validation}
\end{figure}

Fig. \ref{fig:withoutNoise} shows the training accuracy of the model over number of epochs for the turbo encoding and decoding performance without any noise (Problem 1 and 2). In case of encoding, even though the model shows a high training accuracy, the validation accuracy stays constant at 67\% after 3 epochs. On the contrary, the model performs well for decoding the turbo codes with training and validation accuracies approaching 100\% after 6 epochs. This shows us, that the proposed model is good for the decoding operation and not very well suited for encoding.

\begin{figure*}[t!]
	\centering
	\begin{subfigure}[b]{\textwidth}
		\centering
		\includegraphics[width=0.90\linewidth]{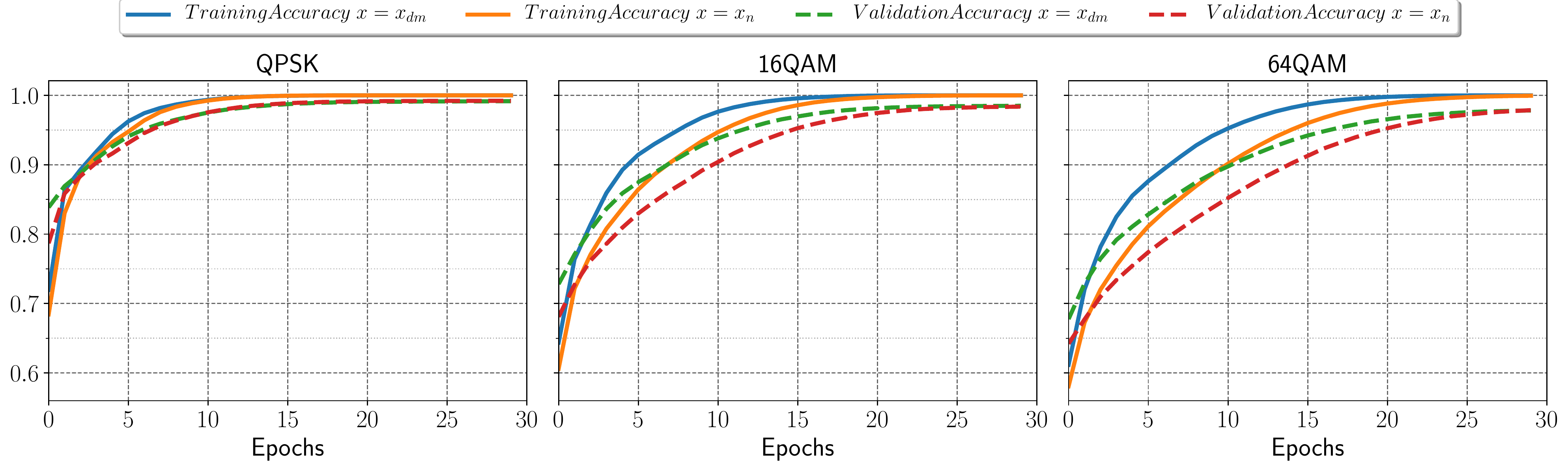}
		\caption{Training and Validation Accuracies}
		\label{fig:Ng3} 
	\end{subfigure}
	
	\begin{subfigure}[b]{\textwidth}
		\centering
		\includegraphics[width=0.90\linewidth]{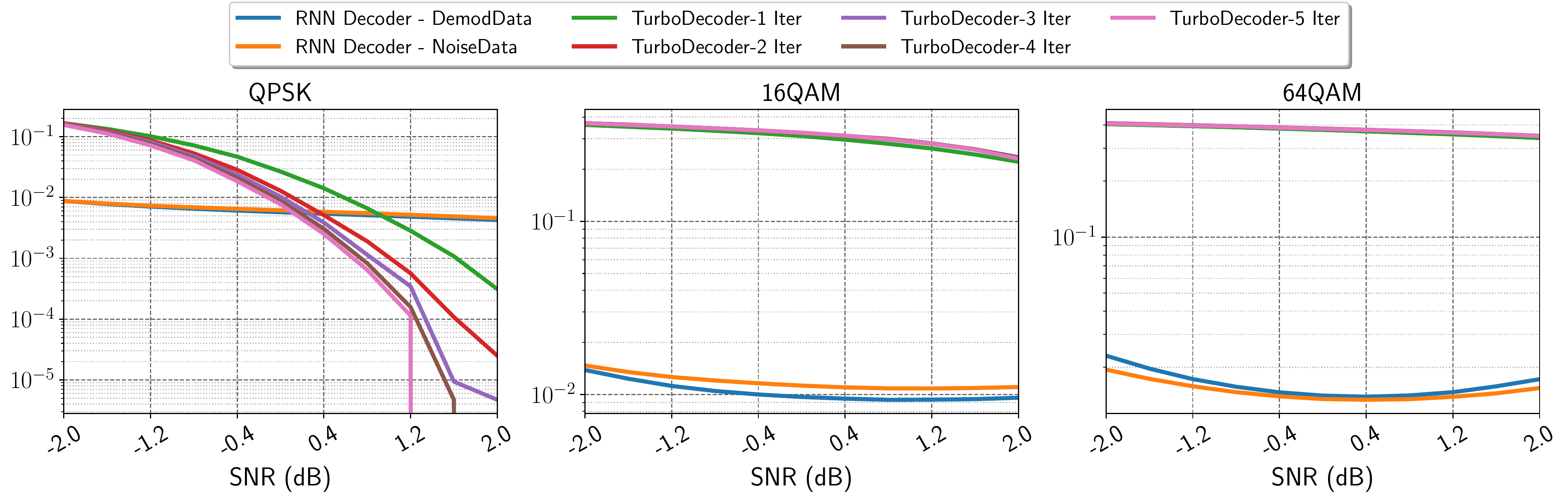}
		\caption{\ac{ber} Curves }
		\label{fig:Ng4}
	\end{subfigure}
	\caption{Decoding Performance - Problem 3 \& 4}
\end{figure*}

The testing accuracies for both the turbo encoding and decoding operations are consistent with the validation accuracies and are outlined in Table \ref{tab:testingAccuracies}

\begin{table}[htb!]
	\centering
	\caption{Testing Accuracy - No Noise}
	\label{tab:testingAccuracies}
	\begin{tabular}{@{}ll@{}}
		\toprule
		Problem            & Testing accuracy \\ \midrule
		Turbo Encoding     & 67\%             \\
		Turbo Decoding - 1 & 100\%            \\ \bottomrule
	\end{tabular}
\end{table}

Figs. \ref{fig:Ng1} and \ref{fig:Ng2} show the effect of number of \ac{gru} units and the number of samples on training and validation accuracies respectively. It can be seen, that if the number of \ac{gru} units is less than 200, the model fails to converge for the selected number of epochs. Hence, the choice of having 800 units is a safe assumption. Similarly, the choice of using 3000 samples for training the model also seems safe given that there is some jitter in accuracy for the number of samples<2000.

\subsection{Turbo Decoding on Demodulated and Channel Output Data}
%\begin{figure*}[t!]
%	\centering 
%	\includegraphics[width=\textwidth]{turboDecodingFinal.pdf}
%	\caption{Demodulated Soft Bits - \ac{ber}}
%	\label{fig:demod}
%\end{figure*} 

For problems 3 and 4, the data from the demodulator and the channel are used respectively. Fig. \ref{fig:Ng3} shows the evolution of training and validation accuracies with respect to the number of epochs for the three modulation schemes. It can be seen clearly that the model shows good training and validation performance on \ac{qpsk} when compared to that of 16-\ac{qam} and 64-\ac{qam}. It can also be seen that the model performance is similar on both the demodulated and noise data which shows the ability of the model to understand the modulation structure thereby eliminating the need for demodulating the data beforehand. A careful look at the validation accuracies when using both the demodulated and noise data reveals, that the model fails to converge for the considered number of epochs.\

Fig. \ref{fig:Ng4} shows the \ac{ber} performance of the model on both the datasets (calculated as $1-\text{Testing Accuracy}$) for each given \ac{snr}. It can be seen that the \ac{rnn} model outperforms the conventional turbo decoder (for all decoding iterations) for low \acp{snr} (<0.4 dB). However, at higher \acp{snr}, the turbo decoder's \ac{ber} drops down exponentially while the drop is only linear for the \ac{rnn} decoder. 

\section{Conclusions \& Future Work}
Driven by the demand for a paradigm shift in communication system design, \ac{ml} is extensively studied by researchers to assess its applicability to wireless networks especially at the \ac{phy} layer. In this regard, this paper presents a \ac{dl} approach for encoding and decoding of turbo codes (LTE variant) using auto-encoders based on \acp{rnn}. To this end, this paper presents an \ac{rnn} architecture based on \acp{gru} and compares its performance with the conventional turbo encoder/decoder. Simulation results show that the proposed \ac{rnn} architecture is able to learn the structure of turbo codes and even outperform the conventional turbo decoder in terms of \ac{ber} at lower \acp{snr}. However, it lags behind the conventional turbo decoder at high \acp{snr} by at least 2 orders of magnitude. A hybrid approach using \ac{rnn} decoder at low \acp{snr} and turbo decoder at high \acp{snr} will be able to provide a performance enhancement in terms of decoding accuracy in the current wireless systems.\

The current analysis considers a fixed data packet size of 64 bits. Without loss of generality, the proposed \ac{rnn} architecture can be extended to support variable sized inputs and outputs. In this case, the problem needs to be reformulated as a sequence-to-sequence learning problem on the lines of natural language translation.

\bibliographystyle{IEEEtran}
\bibliography{RadioML}

\end{document}